\documentclass[aps,pra,floatfix,superscriptaddress,twocolumn,showpacs,10pt]{revtex4-1}
\usepackage{amssymb, amsmath,color,mciteplus,graphicx,subfigure}
\usepackage{calc,epsfig,epstopdf,color,mciteplus,bm,mathrsfs}
\usepackage{times}

\newcommand{\ket}[1]{\vert #1 \rangle}
\newcommand{\bra}[1]{\langle #1 \vert}

\begin{document}

\title{Nonadiabatic holonomic quantum computation on coupled transmons with ancillaries}

\author{Tao Chen}
\affiliation{Guangdong Provincial Key Laboratory of Quantum Engineering and Quantum Materials,  and School of Physics\\ and Telecommunication Engineering, South China Normal University, Guangzhou 510006, China}

\author{Jiang Zhang}
\affiliation{State Key Laboratory of Low-Dimensional Quantum Physics and Department of Physics, Tsinghua University, Beijing 100084, China}

\author{Zheng-Yuan Xue}\email{zyxue83@163.com}
\affiliation{Guangdong Provincial Key Laboratory of Quantum Engineering and Quantum Materials,  and School of Physics\\ and Telecommunication Engineering, South China Normal University, Guangzhou 510006, China}

\date{\today}

\begin{abstract}
  The physical implementation of holonomic quantum computation is challenging due to the needed complex controllable interactions in multilevel quantum systems. Here we propose to implement nonadiabatic holonomic quantum computation with conventional capacitive coupled superconducting transmon qubits. A universal set of holonomic gates is constructed with the help of the interaction with an auxiliary qubit rather than relying on delicate control over an auxiliary level of multilevel quantum systems. Explicitly, these quantum gates are realized by tunable interactions in an all-resonant way, which leads to high-fidelity gate operations. In this way, the distinct merit of our scheme is that we use only the lowest two levels of a transmon to form the qubit states. In addition, the auxiliary qubits are in their ground states before and after every gate operation, so that the holonomic gates can be constructed successively. Therefore, our scheme provides a promising method towards the practical realization of high-fidelity nonadiabatic holonomic quantum computation.
\end{abstract}

\maketitle

\section{introduction}

Quantum computers are believed to outperform their classical counterparts in solving certain hard problems \cite{shor}. However, quantum states are susceptible to noises induced by their surrounding environment, thus the practical implementation of quantum computers is harsh. Since geometric phases depend only on the global properties of the evolution paths, they can effectively resist the influence of certain local noises and thus become a promising medium for quantum computation. Holonomic quantum computation (HQC) \cite{HQC1} is a strategy to build a universal set of robust gates using non-Abelian geometric phases \cite{non-Abelian}. This idea was originally proposed based on adiabatic evolution \cite{Duan,HQC2,HQC3,HQC4,HQC5}, which aims to achieve high-fidelity quantum computation. However, the adiabatic condition requires a long evolution time, during which environmental noises will ruin designed operations.

For practical quantum computation, nonadiabatic evolution is necessary \cite{ZSL,ZSL1}. Recently, nonadiabatic HQC by using the cyclic evolution of a subspace existing in the general three-level $\Lambda$ quantum system has been proposed, so that a universal set of fast geometric quantum gates can be implemented \cite{Tong,XGF,3level1,3level2,3level3,3level4,3level5,3level6,3level7, 3level8,3level9,3level10, 3level11,3level12}. This type of nonadiabatic gates has been experimentally demonstrated in superconducting circuits \cite{Abdumalikov2013,ibmexp,xuy,sust}, NMR \cite{Feng2013,li2017}, and  electron spins in diamond \cite{Zu2014, Arroyo-Camejo2014,nv2017,nv20172}.

The existing nonadiabatic HQC schemes based on three-level systems require the use of a third auxiliary energy level. This is challenging for a superconducting transmon, due to the fact that the corresponding energy spectrum is only weakly anharmonic, which makes implementation of the controllable interactions between qubits difficult. In addition, problems will occur when applying these schemes to quantum error correction. Since the required projective measurement on multilevel quantum systems can have a state collapse to the auxiliary level, rather than the levels used for a qubit. In order to avoid this drawback, we propose to realize a nonadiabatic HQC scheme \cite{fourlevel1, fourlevel2, fourlevel3, fourlevel4} with capacitive coupled superconducting transmon qubits, where we construct universal holonomic gates with the help of auxiliary qubits rather than auxiliary levels. The distinct merit of our scheme is that the auxiliary qubits are in their ground states before and after each gate operation so that the problems caused by auxiliary levels can be overcome. In addition, our scheme uses the lowest two levels of a transmon to form the qubit states and can result in universal HQC with conventional resonant interactions, leading to fast and high-fidelity universal quantum gates. Moreover, to obtain tunable all-resonant interactions between the target and the auxiliary qubits, we only need to add modulations on the target qubits by well-controlled microwaves. Therefore, our scheme can be readily implemented in a two-dimensional (2D) lattice composed of coupled superconducting transmons and, thus, offers promising scalability.

\section{Arbitrary single-qubit gate}

\begin{figure}
  \centering
\includegraphics[width=7cm]{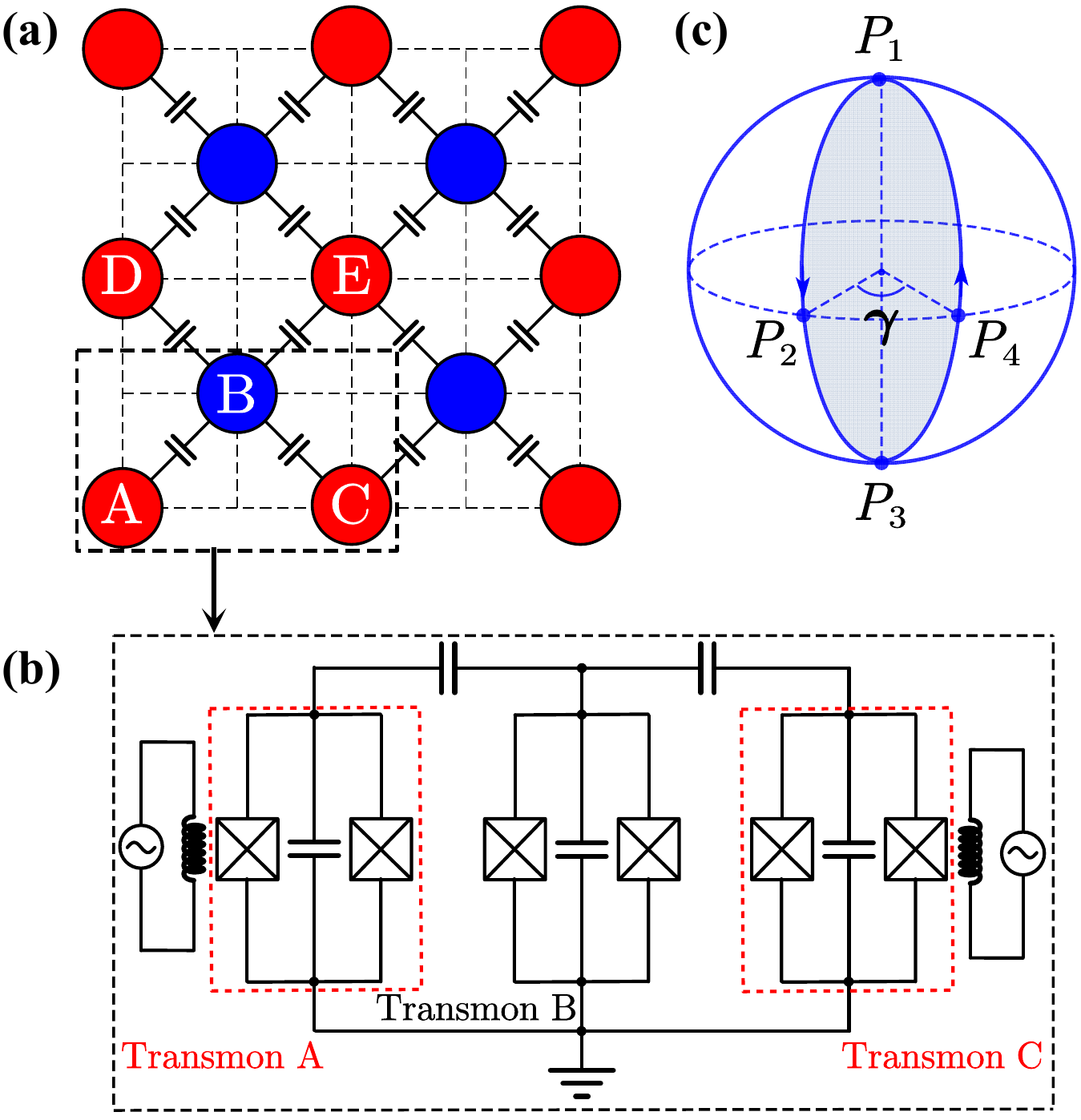}
  \caption{Proposed setup of our scheme. (a) Configuration of the 2D lattice consisting of coupled superconducting transmons. Filled red circles represent target transmon qubits; filled blue circles represent auxiliary transmon qubits. (b) Transmon A/C interacts with transmon B via the capacitance between them, the tunable interaction of which can be obtained by adding a periodical modulation to the transition frequency of transmon A/C. (c) Orange-slice-shaped evolution path in the Bloch sphere.}
  \label{Fig1}
\end{figure}

The setup we consider is a 2D lattice composed of coupled superconducting transmons \cite{transmon} with different frequencies, as illustrated in Fig.~\ref{Fig1}(a). The lowest two levels of each transmon are used to define a qubit. There are two kinds of qubits in the lattice: those storing quantum information (called target qubits) and those used to assist in constructing quantum gates on the target qubits (called auxiliary qubits). Explicitly, single-qubit holonomic gates on a target qubit (e.g., qubit A) can be built with the help of a nearby auxiliary qubit (e.g., qubit B); an entangling two-qubit holonomic gate on a pair of adjacent target qubits (e.g., qubits A and C or D) can be realized by an auxiliary qubit (e.g., qubit B) connecting to both of them. Moreover, through qubit B, a two-qubit gate on qubits A and E can also be constructed directly, offering a more efficient scheme to manipulate two remote target qubits. This indicates the scalability of our proposal.

In order to realize an arbitrary single-qubit holonomic gate on a target transmon qubit (e.g., qubit A), we introduce auxiliary transmon qubit B, which is driven by a classical field and capacitively coupled to qubit A, as shown in Fig. \ref{Fig1}(b). Furthermore, we consider a realistic case of the transmon, i.e., it is weakly anharmonic, so that we need to take the third energy level into account, since the main leakage out of the qubit basis comes from this level. Assuming that $\hbar=1$ hereafter, the Hamiltonian of the coupled system can be expressed as
\begin{eqnarray}
\label{Eq1}
H_{c1}&=&\sum_{\chi=\mathrm{A},\mathrm{B}} \omega_\chi n_\chi+\frac{\alpha_\chi}{2}(1-n_\chi)n_\chi  \notag \\
&+& g_{_{\mathrm{AB}}}(a^\dagger b+a b^\dagger)+\varepsilon \cos(\omega t-\phi)(b^\dagger+b),
\end{eqnarray}
where $n_{_{\mathrm{A}}}=a^\dagger a$ and $n_{_{\mathrm{B}}}=b^\dagger b$, with $a=|0\rangle_{_{\mathrm{A}}} \langle 1|+\sqrt{2}|1\rangle_{_{\mathrm{A}}} \langle 2|$ and $b=|0\rangle_{_{\mathrm{B}}} \langle 1|+\sqrt{2}|1\rangle_{_{\mathrm{B}}} \langle 2|$ being the standard lower operators for transmons A and B, respectively; $\omega_{_{\mathrm{A}}}$ and $\omega_{_{\mathrm{B}}}$ are the associated transition frequencies with $\alpha_{_{\mathrm{A}}}$ and $\alpha_{_{\mathrm{B}}}$ being the intrinsic anharmonicities of transmons A and B, respectively; $g_{_{\mathrm{AB}}}$ is the transmon-transmon coupling strength; and $\varepsilon$, $\omega$, and $\phi$ are the classical driving strength, frequency, and phase of transmon B, respectively.

Since two transmons usually do not have identical energy splitting, we add a driving on transmon A to induce effectively resonant coupling between them \cite{flux-driven1,flux-driven0,flux-driven2,flux-driven3,lix}. This can be realized by adding an ac magnetic flux through the loop of transmon A. In this way, a periodical modulation of transmon A's transition frequency of
\begin{eqnarray}
\label{Eq2}
\omega_{_{\mathrm{A}}}(t)=\omega_{_{\mathrm{A}}}+\epsilon_1 \sin(\nu_1 t+\pi/2)
\end{eqnarray}
can be achieved. Moving into the interaction picture, the interaction Hamiltonian becomes
\begin{eqnarray}
\label{Eq3}
H_I&=& g_{_{\mathrm{AB}}} \left\{ |10\rangle\langle 01|e^{-i\Delta_1 t}e^{-i\beta_1 \cos(\nu_1 t+\pi/2)}\right.                   \notag \\
&+&\left.\sqrt{2}|11\rangle\langle 02|e^{-i(\Delta_1-\alpha_{_{\mathrm{B}}}) t}e^{-i\beta_1 \cos(\nu_1 t+\pi/2)}\right.                  \notag \\
&+&\left.\sqrt{2}|20\rangle\langle 11|e^{-i(\Delta_1+\alpha_{_{\mathrm{A}}}) t}e^{-i\beta_1 \cos(\nu_1 t+\pi/2)} \right\}                \notag \\
&+&\frac{\varepsilon}{2}\sum_{j=0,1} \sqrt{j+1} |j+1\rangle_{_{\mathrm{B}}}\langle j|e^{i(\delta-j\alpha_{_{\mathrm{B}}}) t}e^{i\phi}+ \mathrm{H.c.},
\end{eqnarray}
where $\Delta_1=\omega_{_{\mathrm{B}}}-\omega_{_{\mathrm{A}}}$, $\delta=\omega_{_{\mathrm{B}}}-\omega$, $\beta_1=\epsilon_1/\nu_1$, and $|mn\rangle=|m\rangle_{_{\mathrm{A}}}\otimes|n\rangle_{_{\mathrm{B}}}$. We consider the case of resonant driving on transmon qubit B ($\delta=0$), and parametric driving compensates the energy difference between transmon qubit A and transmon qubit B, i.e., $\Delta_1=\nu_1$. Then, using the Jacobi-Anger identity,
\begin{eqnarray}
\label{Eq0}
e^{-i\beta_1 \cos(\nu_1 t+\pi/2)}=\sum^{\infty}_{m=-\infty}(-i)^m J_m(\beta_1) e^{-im(\nu_1 t+\pi/2)}, \notag
\end{eqnarray}
where $J_m(\beta_1)$ is a Bessel function of the first kind. Finally, applying the rotating-wave approximation, we obtain the effective resonant interaction Hamiltonian as
\begin{eqnarray}
\label{Eq4}
H_1= g^{\prime}_{_{\mathrm{AB}}} |10\rangle\langle 01|+\frac{\varepsilon}{2}|1\rangle_{_{\mathrm{B}}}\langle 0|e^{i\phi}+ \mathrm{H.c.},
\end{eqnarray}
where $g^{\prime}_{_{\mathrm{AB}}}=J_1(\beta_1)g_{_{\mathrm{AB}}}$.

By setting $g^{\prime}_{_{\mathrm{AB}}}=\Omega\cos(\theta/2)$, $\varepsilon=\Omega\sin(\theta/2)$ with $\Omega=\sqrt{g^{\prime 2}_{_{\mathrm{AB}}}+\varepsilon^2}$, and $\theta=2\tan^{-1}(\varepsilon/g^{\prime}_{_{\mathrm{AB}}})$, the Hamiltonian in Eq. (\ref{Eq4}) reduces to the block off-diagonal form
\begin{eqnarray}
\label{Eq5}
H_1=
{\Omega} \left(
\begin{array}{cc}
0 & F \\
F^\dagger & 0 \\
\end{array}
\right),
\end{eqnarray}
in the basis $S=\{|00\rangle,|10\rangle,|01\rangle,|11\rangle\}$ with
\begin{eqnarray}
\label{Eq6}
F=
\left(
\begin{array}{cc}
\frac{1}{2} \sin\frac{\theta}{2} e^{-i\phi} & 0 \\
\cos\frac{\theta}{2} & \frac{1}{2} \sin\frac{\theta}{2} e^{-i\phi} \\
\end{array}
\right).
\end{eqnarray}
Since matrix $F$ is invertible, it has a unique singular value decomposition $F=WQ R^\dagger$, where
\begin{eqnarray}
\label{Eq7}
W&=&
\left(
\begin{array}{cc}
\sin\frac{\theta}{4} & \cos\frac{\theta}{4}e^{-i\phi} \\
 \cos\frac{\theta}{4}e^{i\phi} & -\sin\frac{\theta}{4} \notag\\
\end{array}
\right),\notag\\
Q&=&
\left(
\begin{array}{cc}
\cos^2\frac{\theta}{4} & 0 \\
0 & \sin^2\frac{\theta}{4} \\
\end{array}
\right),\notag\\
R^\dagger&=&
\left(
\begin{array}{cc}
\cos\frac{\theta}{4} & \sin\frac{\theta}{4}e^{-i\phi} \\
\sin\frac{\theta}{4}e^{i\phi} & -\cos\frac{\theta}{4} \\
\end{array}
\right).
\end{eqnarray}

We can separate the four-dimensional (4D) Hilbert space of the Hamiltonian $H_1$ into two 2D subspaces of
\begin{eqnarray}
\label{Eq8}
S=S_0\oplus S_1,
\end{eqnarray}
where $S_0=\text{Span}\{|00\rangle,|10\rangle\}$ and $S_1=\text{Span}\{|01\rangle,|11\rangle\}$. This implies that in the basis $S$ the evolution operator splits into $2\times2$ blocks, which can be expressed as
\begin{eqnarray}
\label{Eq9}
U_1(\tau)=
\left(
\begin{array}{cc}
W\cos(a_\tau Q)W^\dagger & -iW\sin(a_\tau Q)R^\dagger \\
-iR\sin(a_\tau Q)W^\dagger & R\cos(a_\tau Q)R^\dagger \\
\end{array}
\right),
\end{eqnarray}
where $a_\tau=\Omega \tau$, with $\tau$ being the total evolution time.

The 4D model can be used to achieve an arbitrary single-qubit holonomic gate on target qubit A. The corresponding gate construction is realized by evolving the system along an orange-slice-shaped path \cite{OSSE1,fourlevel2,OSSE2}, as shown in Fig. \ref{Fig1}(c). In the first path segment $[0,\frac{\tau}{2}]$, we set $\phi=0$ and obtain $F_1=W_1Q_1R^\dagger_1$. Subspace $S_0$ is evolved into subspace $S_1$ via the path $P_1\rightarrow P_2\rightarrow P_3$ by choosing a certain time such that $\cos(a_{\frac{\tau}{2}} Q_1)=0$ and $\sin(a_{\frac{\tau}{2}} Q_1)=G_i$ , where $i=I$ or $z$, $G_I=\text{diag}\{1,1\}$, $G_z=\text{diag}\{1,-1\}$. We note that, in order to satisfy the above conditions, one needs to make sure that the parameter $\theta\neq 2n\pi$, where $n$ is an integer. At point $P_3$ in the Bloch sphere, we change the relative phase parameter $\phi=\gamma$. Similarly, we define $F_2=W_2Q_2R^\dagger_2$ in the second path segment $[\frac{\tau}{2},\tau]$ and then return $S_1$ to the initial subspace $S_0$ via path $P_3\rightarrow P_4\rightarrow P_1$ by choosing a certain time such that $\cos(a_{\frac{\tau}{2}}Q_2)=0$, $\sin(a_{\frac{\tau}{2}}Q_2)=G_i$. We obtain the final evolution operator as
\begin{eqnarray}
\label{Eq10}
U_1(\tau)&=&U\left(\tau, \frac{\tau}{2}\right)U\left(\frac{\tau}{2}, 0\right) \notag \\
&=& U_0 \otimes|0\rangle_{_{\mathrm{B}}}\langle0| + U_1 \otimes|1\rangle_{_{\mathrm{B}}}\langle1|,
\end{eqnarray}
where the evolution operators $U_0=-W_2G_i R^\dagger_2R_1G_iW^\dagger_1$ and $U_1=-R_2G_iW^\dagger_2W_1G_i R^\dagger_1$ act on target qubit A conditionalized on states $|0\rangle_{_{\mathrm{B}}}$ and $|1\rangle_{_{\mathrm{B}}}$ of auxiliary transmon qubit B, respectively.
\begin{figure}[tbp]
  \centering
\includegraphics[width=8cm]{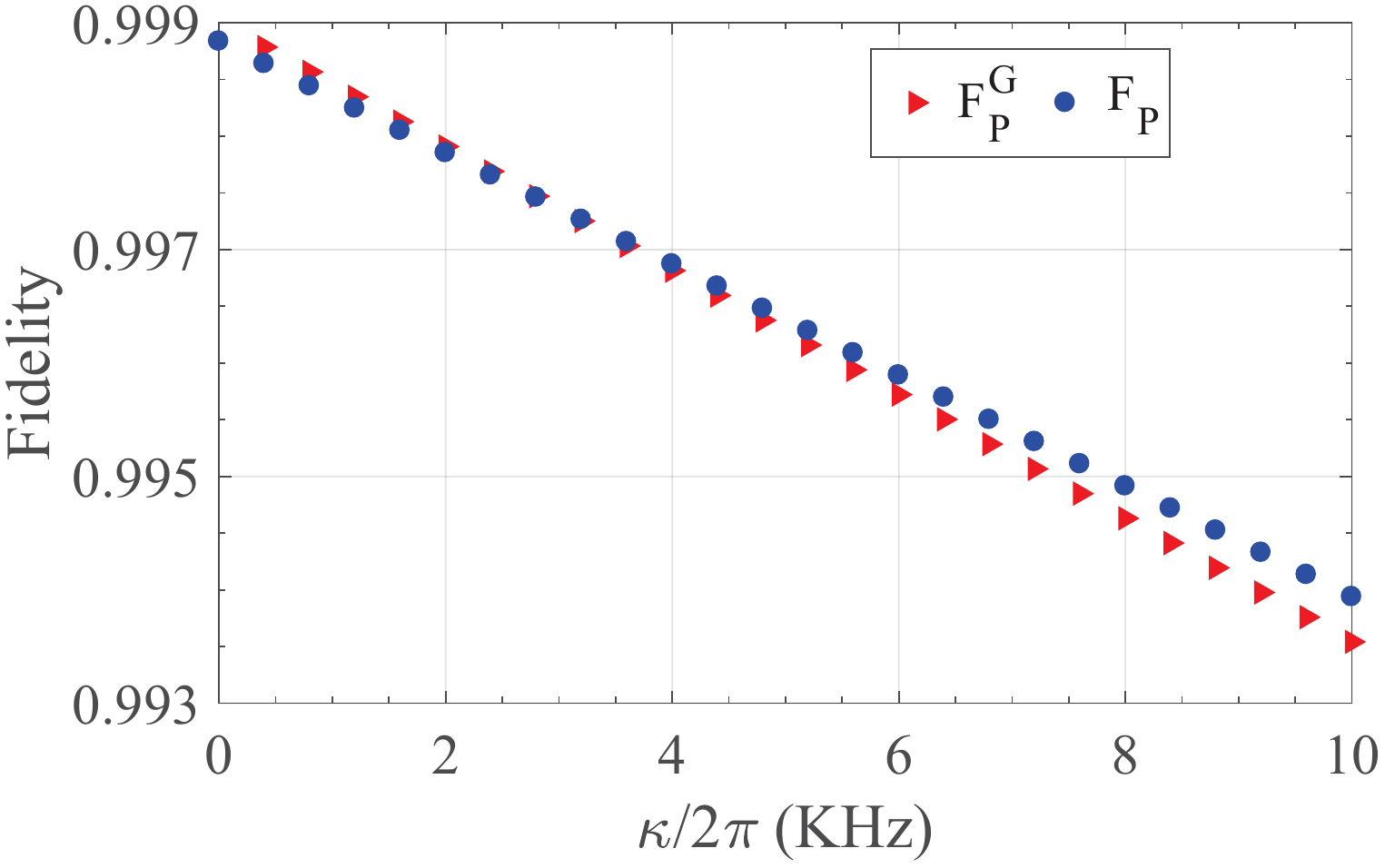}
  \caption{Simulation results for the $U_P$ gate. State fidelity $F_P$ of the initial state $\frac{1}{\sqrt2}(|00\rangle+|10\rangle)$ and gate fidelity $F^G_P$ as a function of the different decoherence rate.}
  \label{Fig2}
\end{figure}
If we initially prepare auxiliary qubit B in its ground state $|0\rangle_{_{\mathrm{B}}}$, we select the evolution operator $U_0$ in the orthogonal subspace $S_0=\{|00\rangle,|10\rangle\}$. We further set $G_i=G_I$ and $\gamma=\pi$; the related evolution operator on qubit A reads
\begin{eqnarray}
\label{Eq11}
U^\prime_0=
\left(
\begin{array}{cc}
 \cos\theta & -\sin\theta \\
\sin\theta & \cos\theta \\
\end{array}
\right)
=e^{-i\theta \sigma_y},
\end{eqnarray}
which is a rotation around the \emph{Y} axis by an angle $\theta$. Alternatively, if we take $G_i=G_z$, the corresponding evolution operator becomes
\begin{eqnarray}
\label{Eq12}
U^{\prime\prime}_0=
\left(
\begin{array}{cc}
 e^{-i \gamma} & 0 \\
0 & e^{i \gamma} \\
\end{array}
\right)=e^{-i\gamma \sigma_z},
\end{eqnarray}
which is a rotation around the \emph{Z} axis by an angle $\gamma$.

The above processes are nonadiabatic holonomy transformations, because the following two conditions are satisfied: (i) Hamiltonian $H_1$ vanishes in the evolving subspace $S_0$, i.e.,
\begin{eqnarray}
\label{Eq13}
P_0H_1P_0=U_1(t)P_0H_1P_0U^\dagger_1(t)=0,
\end{eqnarray}
where $P_0=|00\rangle\langle00|+|10\rangle\langle10|$. It shows that the evolution satisfies the parallel-transport condition.
(ii) The evolution of the subspace $S_0$ is cyclic, since
\begin{eqnarray}
\label{Eq14}
S_0(\tau) &\equiv& \text{Span}\{U_1(\tau)|00\rangle,U_1(\tau)|10\rangle\} \notag \\
&=&\text{Span}\{|00\rangle,|10\rangle\}=S_0.
\end{eqnarray}
Thus, an arbitrary single-qubit holonomic gate on target qubit A can be implemented.

However, in practical physical implementations, decoherence is unavoidable. Therefore, we consider the decoherence effect by numerical simulating the Lindblad master equation,
\begin{eqnarray}
\label{Eq15}
\dot\rho_1&=& -i[H_{c1}, \rho_1]+\sum_{\chi=\mathrm{A},\mathrm{B}}\left\{ \frac {\kappa^\chi_-} {2}\mathscr{L}(|0\rangle_\chi \langle 1|+2|1\rangle_\chi \langle 2|)\right. \notag \\
&& \left. + \frac {\kappa^\chi_z} {2}\mathscr{L}(|1\rangle_\chi \langle 1|+2|2\rangle_\chi \langle 2|) \right\},
\end{eqnarray}
where $\rho_1$ is the density matrix of the considered system, $\mathscr{L}(\mathcal{A})=2\mathcal{A}\rho_1
\mathcal{A}^\dagger-\mathcal{A}^\dagger \mathcal{A} \rho_1 -\rho_1 \mathcal{A}^\dagger \mathcal{A}$ for operator $\mathcal{A}$, and $\kappa^\chi_-$ and $\kappa^\chi_z$ are the relaxation and dephasing rates of the transmon, respectively.  According to recent superconducting experiments, we choose $\alpha_{_{\mathrm{A}}}=2\pi\times 375$ MHz, $\alpha_{_{\mathrm{B}}}=2\pi\times 350$ MHz \cite{anharmonicity}, $\Delta_1=2\pi \times 245$ MHz \cite{flux-driven3}, $\beta_1=\epsilon_1/\nu_1=\epsilon_1/\Delta_1\approx 1.6$, $\Omega=2\pi\times 13$ MHz (when $g_{_{\mathrm{AB}}}\approx2\pi\times 11.41$ MHz and $\varepsilon\approx2\pi\times 11.26$ MHz). As demonstrated in Ref. \cite{decoherence}, the relaxation and dephasing rates of a transmon qubit are of the same order, about $2\pi\times 1.5$ kHz. Here, we assume that the relaxation and dephasing rates of the transmon are the same, i.e., $\kappa^\mathrm{A}_-=\kappa^\mathrm{A}_z=\kappa^\mathrm{B}_-=\kappa^\mathrm{B}_z=\kappa\in2\pi\times[0,10]$ kHz. We take the phase-shift gate $U_P=\text{diag}\{1,e^{i\frac{\pi}{4}}\}$ as an example, which corresponds to $\gamma=\pi/8$ in Eq. (\ref{Eq12}). Assuming that the initial state of the two qubits is $\frac{1}{\sqrt2}(|00\rangle+|10\rangle)$, the phase-shift gate results in the ideal final state $|\psi_{f_P}\rangle=\frac{1}{\sqrt2}(|00\rangle+e^{i\frac{\pi}{4}}|10\rangle)$. We first evaluate the performance of this gate by the state fidelity defined by $F_P=\langle\psi_{f_P}|\rho_1|\psi_{f_P}\rangle$. In Fig. \ref{Fig2}, we use blue circles to show the trend of the state fidelity with the decoherence rate $\kappa$, and find that the state fidelity can reach about $99.64\%$ for $\kappa=2\pi \times5$ kHz. Moreover, to fully evaluate the gate, for a general initial state $|\psi_{1}\rangle=\cos\theta_1|00\rangle+\sin\theta_1|10\rangle$, $U_P$ results in an ideal final state $|\psi_{f}\rangle=\cos\theta_1|00\rangle+ e^{i\frac{\pi}{4}}\sin\theta_1|10\rangle$; we define the gate fidelity as $F^G_P=\frac {1} {2\pi}\int_0^{2\pi} \langle  \psi_{f}|\rho_1|\psi_{f}\rangle d\theta_1$ \cite{gatefidelity1}, with the integration numerically done for 1001 input states with $\theta_1$ uniformly distributed over $[0, 2\pi]$. In Fig. \ref{Fig2}, we plot the gate fidelity as a function of the decoherence rate $\kappa$ with red triangles. We find that the gate fidelity is also about $99.63\%$ for $\kappa=2\pi \times5$ kHz, which is experimentally accessible. Finally, we want to emphasize that all the simulation hereafter is based on the original interaction Hamiltonian in Eq. (\ref{Eq1}) without any approximation, thus verifying our analytical results.

\begin{figure}[tbp]
  \centering
\includegraphics[width=8cm]{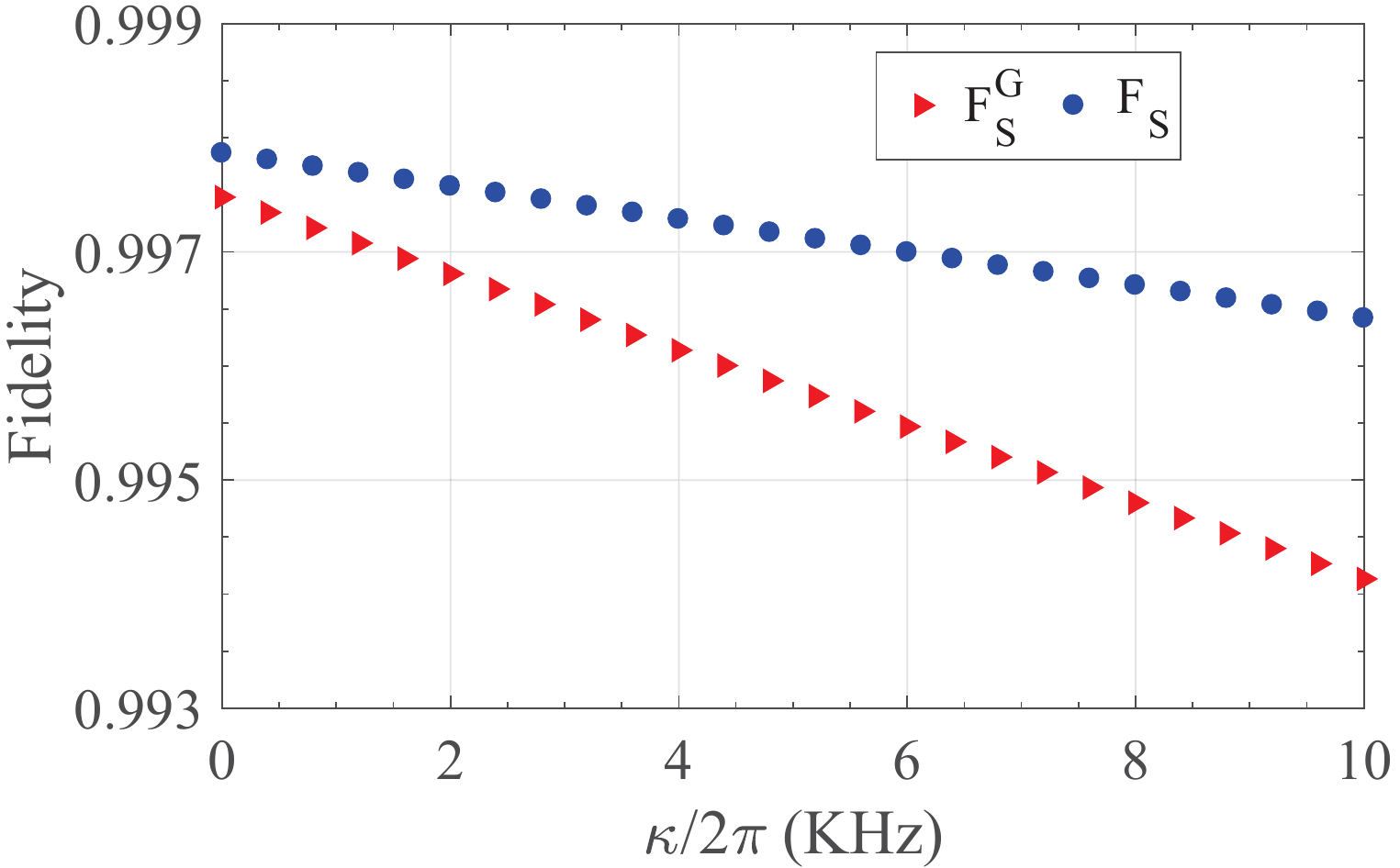}
  \caption{Simulation results for the SWAP-like two-qubit gate. State fidelity $F_S$ of the initial state $\frac{1}{\sqrt2}(|000\rangle+|001\rangle)$ and gate fidelity $F^G_S$ as a function of the decoherence rate.}
  \label{Fig3}
\end{figure}

\section{Nontrivial two-qubit gates}

Next we consider the construction of a kind of nontrivial two-qubit holonomic gate on a pair of target qubits in the lattice. Combining with the single-qubit holonomic gates, nonadiabatic holonomic quantum computation can then be realized. As shown in Fig. \ref{Fig1}(a), we choose two target qubits, A and C, to capacitively couple to the same auxiliary qubit B. This three-qubit system is described by
\begin{eqnarray}
\label{Eq16}
H_{c2}&=&\sum_{\chi=\mathrm{A},\mathrm{B},\mathrm{C}} \omega_\chi n_\chi+\frac{\alpha_\chi}{2}(1-n_\chi)n_\chi  \notag \\
&+&g_{_{\mathrm{AB}}}(a^\dagger b+a b^\dagger)+g_{_{\mathrm{BC}}}(b^\dagger c+b c^\dagger).
\end{eqnarray}

In the following, to get the resonant interaction between transmon qubit pair A and B and pair B and C, we separately add two different parametric modulations to the transition frequency of transmons A and C by two microwave fields, which are
\begin{eqnarray}
\label{Eq17}
\omega_{_{\mathrm{A}}}(t)&=&\omega_{_{\mathrm{A}}}+\epsilon_1 \sin(\nu_1 t+\pi/2),  \notag \\
\omega_{_{\mathrm{C}}}(t)&=&\omega_{_{\mathrm{C}}}+\epsilon_2 \sin(\nu_2 t+\varphi+\pi/2).
\end{eqnarray}
By applying the rotating-wave approximation and setting $\Delta_1=\nu_1$ and $\Delta_2=\omega_{_{\mathrm{B}}}-\omega_{_{\mathrm{C}}}=\nu_2$, the finally effective resonant interaction Hamiltonian can be shown as
\begin{eqnarray}
\label{Eq18}
H_2= g^{\prime}_{_{\mathrm{AB}}} |01\rangle_{_{\mathrm{AB}}}\langle 10|+g^{\prime}_{_{\mathrm{BC}}}e^{i \varphi} |01\rangle_{_{\mathrm{BC}}}\langle 10|+ \mathrm{H.c.}, \notag \\
\end{eqnarray}
where $g^{\prime}_{_{\mathrm{AB}}}=J_1(\beta_1)g_{_{\mathrm{AB}}}$, $\beta_1=\epsilon_1/\nu_1$ and $g^{\prime}_{_{\mathrm{BC}}}=J_1(\beta_2)g_{_{\mathrm{BC}}}$, $\beta_2=\epsilon_2/\nu_2$.

Next, we explain how the above effective resonant interaction Hamiltonian can be used to achieve nontrivial two-qubit holonomic gates on target qubits A and C. Resetting $g^{\prime}_{_{\mathrm{AB}}}=g\cos(\vartheta/2)$, $g^{\prime}_{_{\mathrm{BC}}}=g\sin(\vartheta/2)$ with $g=\sqrt{g^{\prime 2}_{_{\mathrm{AB}}}+g^{\prime 2}_{_{\mathrm{BC}}}}$ and $\vartheta=2\tan^{-1}(g^{\prime}_{_{\mathrm{BC}}}/g^{\prime}_{_{\mathrm{AB}}})$, the Hamiltonian can be reduced to the block off-diagonal form
\begin{eqnarray}
\label{Eq19}
H_2=g (|0\rangle_{_{\mathrm{B}}}\langle 1|\otimes K+|1\rangle_{_{\mathrm{B}}}\langle 0|\otimes K^\dagger),
\end{eqnarray}
where
\begin{eqnarray}
\label{Eq20}
K=
\left(
\begin{array}{cccc}
 0 & 0 & 0 & 0 \\
 \sin\frac{\vartheta} {2} e^{i \varphi} & 0 & 0 & 0 \\
 \cos\frac{\vartheta} {2} & 0 & 0 & 0 \\
 0 & \cos\frac{\vartheta} {2}& \sin\frac{\vartheta} {2} e^{i \varphi} & 0 \\
\end{array}
\right)
\end{eqnarray}
is within the 4D orthonormal basis $\{|00\rangle_{_{\mathrm{AC}}},|01\rangle_{_{\mathrm{AC}}},|10\rangle_{_{\mathrm{AC}}}, |11\rangle_{_{\mathrm{AC}}}\}$. We perform a unique singular value decomposition on matrix $K$ in the form $K=XYZ^\dag$ with
\begin{eqnarray}
\label{Eq21}
X&=&
\left(
\begin{array}{cccc}
 1 & 0 & 0 & 0 \\
 0 & \cos\frac{\vartheta} {2}  & 0 & \sin\frac{\vartheta} {2} e^{i \varphi} \\
 0 & -\sin\frac{\vartheta} {2} e^{-i \varphi} & 0 & \cos\frac{\vartheta} {2}  \\
 0 & 0 & 1 & 0 \\
\end{array}
\right),  \notag \\
Y&=&
\left(
\begin{array}{cccc}
 0 & 0 & 0 & 0 \\
 0 & 0 & 0 & 0 \\
 0 & 0 & 1 & 0 \\
 0 & 0 & 0 & 1 \\
\end{array}
\right),\notag\\
Z^\dag&=&
\left(
\begin{array}{cccc}
 0 & 0 & 0 & 1 \\
 0 & \sin\frac{\vartheta} {2} e^{-i \varphi} & -\cos\frac{\vartheta} {2} & 0 \\
 0 & \cos\frac{\vartheta} {2}  & \sin\frac{\vartheta} {2} e^{i \varphi} & 0 \\
 1 & 0 & 0 & 0 \\
\end{array}
\right).
\end{eqnarray}
Here, we separate the eight-dimensional Hilbert space of the Hamiltonian $H_2$ into two 4D subspaces, i.e.,
\begin{eqnarray}
\label{Eq22}
M=M_0\oplus M_1,
\end{eqnarray}
where $M_0=\text{Span}\{|000\rangle,|001\rangle,|100\rangle,|101\rangle\}$ and $M_1=\text{Span}\{|010\rangle,|011\rangle,|110\rangle,|111\rangle\}$ ( $|ijk\rangle=|i\rangle_{_{\mathrm{A}}}\otimes|j\rangle_{_{\mathrm{B}}}\otimes|k\rangle_{_{\mathrm{C}}}$). Accordingly, the evolution operator generated by $H_2$ also splits into two $4\times4$ blocks, reading
\begin{eqnarray}
\label{Eq23}
U_2(T)=
\left(
\begin{array}{cc}
X\cos(b_T Y)X^\dagger & -iX\sin(b_T Y)Z^\dagger \\
-iZ\sin(b_T Y)X^\dagger & Z\cos(b_T Y)Z^\dagger \\
\end{array}
\right),\ \
\end{eqnarray}
where $b_T=g T$ with $T$ being the evolution time. By choosing a certain time such that $\cos(b_T Y)=J=\text{diag}\{1,1,-1,-1\}$ and $\sin(b_T Y)=\text{diag}\{0,0,0,0\}$, we can obtain
\begin{eqnarray}
\label{Eq24}
U_2(T)= V_0 \otimes|0\rangle_{_{\mathrm{B}}}\langle0| + V_1 \otimes|1\rangle_{_{\mathrm{B}}}\langle1|,
\end{eqnarray}
where the evolution operator $V_0=XJX^\dagger$ and $V_1=ZJZ^\dagger$ act on target qubits A and C, respectively.

The obtained evolution operator is of a holonomic nature, because the following two conditions can be satisfied: (i) Due to $[H_2,U_2(t)]=0$, Hamiltonian $H_2$ vanishes in evolving subspace $M_0$, i.e.,
\begin{eqnarray}
\label{Eq25}
L_0H_2L_0=U_2(t)L_0H_2L_0U^\dagger_2(t)=0,
\end{eqnarray}
where $L_0=|000\rangle\langle000|+|001\rangle\langle001|+|100\rangle\langle100|+|101\rangle\langle101|$;
and (ii) the evolution of the orthogonal subspaces $M_0$ undergoes cyclic evolution, since
\begin{eqnarray}
\label{Eq26}
M_0(T) &\equiv& \text{Span}\{U_2(T)|000\rangle,U_2(T)|001\rangle,        \notag \\
&&U_2(T)|100\rangle,U_2(T)|101\rangle\} \notag \\
&=&\text{Span}\{|000\rangle,|001\rangle,|100\rangle,|101\rangle\}=M_0.
\end{eqnarray}
A similar discussion is valid for the other subspace $M_1$.

Then, by initially preparing auxiliary qubit B in $|0\rangle_{_{\mathrm{B}}}$, i.e., selecting the evolution operator $V_0$ in subspace $M_0=\text{Span}\{|000\rangle,|001\rangle,|100\rangle,|101\rangle\}$, we can obtain a nontrivial two-qubit holonomic gate,
\begin{eqnarray}
\label{Eq27}
V_0&=&X \times \text{diag}\{1,1,-1,-1\}  \times X^\dagger     \notag \\
&=&
\left(
\begin{array}{cccc}
 1 & 0 & 0 & 0 \\
 0 & \cos\vartheta & -\sin\vartheta e^{i \varphi} & 0 \\
 0 & -\sin\vartheta e^{-i\varphi} & -\cos\vartheta & 0 \\
 0 & 0 & 0 & -1 \\
\end{array}
\right),
\end{eqnarray}
on target qubits A and C. When $\varphi=\pi$ and $\vartheta=\pi/2$, we can get a SWAP-like two-qubit gate \cite{SWAPlike},
\begin{eqnarray}
\label{Eq28}
V_S=
\left(
\begin{array}{cccc}
 1 & 0 & 0 & 0 \\
 0 & 0 & 1 & 0 \\
 0 & 1 & 0 & 0 \\
 0 & 0 & 0 & -1 \\
\end{array}
\right),
\end{eqnarray}
which is a nontrivial two-qubit gate for quantum computation.

\begin{figure}[tbp]
  \centering
\includegraphics[width=8cm]{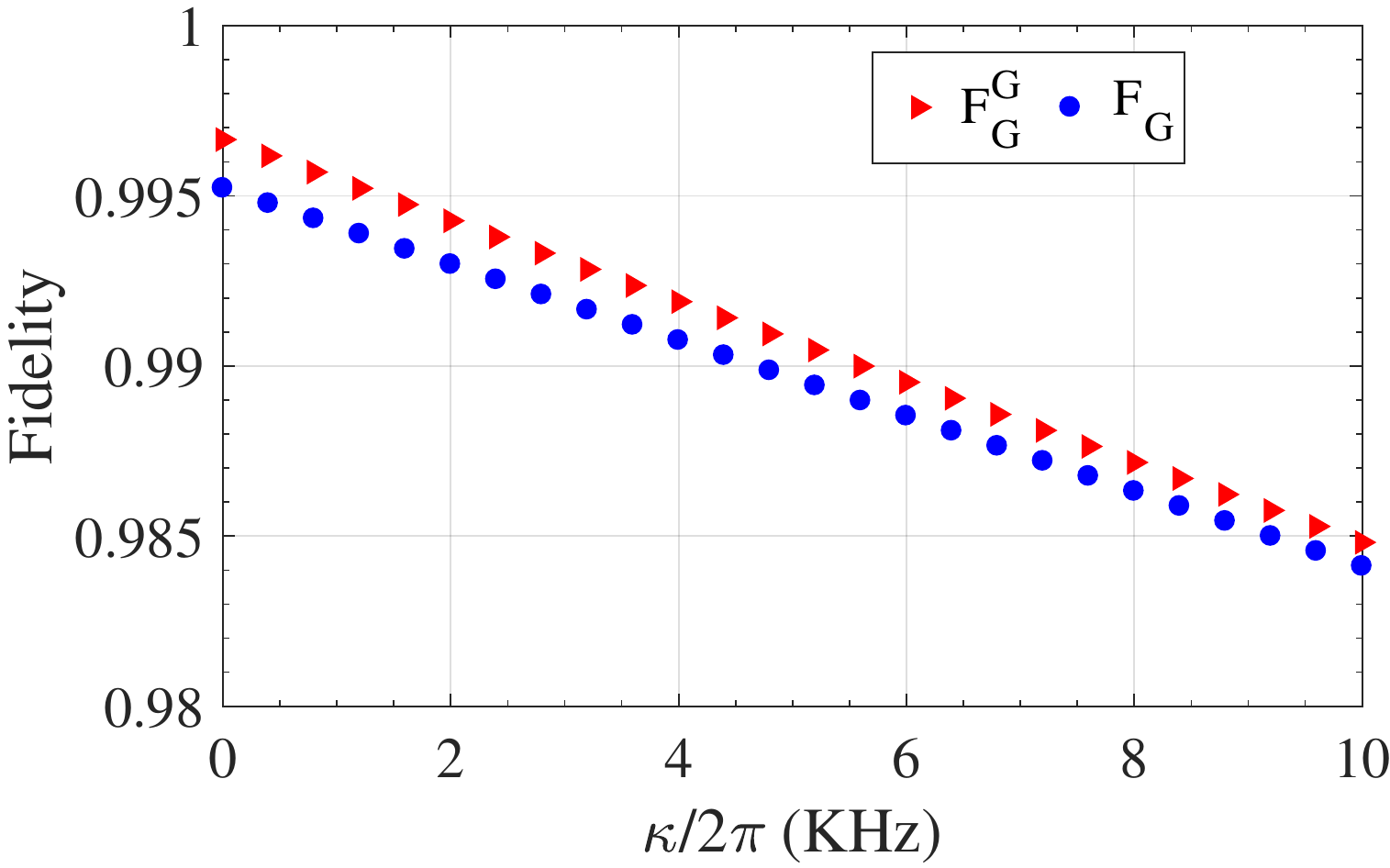}
  \caption{State fidelity $F_G$ of the gate sequence with the initial state being $\frac{1}{\sqrt2}(|000\rangle+|001\rangle)$ and gate fidelity $F^G_G$ as a function of the decoherence rate.}
  \label{Fig4}
\end{figure}

We next take the above SWAP-like gate as an example to verify the performance of this kind of nontrivial two-qubit gate. Here we choose $\Delta_1=2\pi \times245$ MHz, $\Delta_2=2\pi \times230$ MHz, and $\beta_1=\beta_2\approx1.6$. We set the value of anharmonicity $\alpha_{_{\mathrm{C}}}=2\pi\times310$ MHz and the transmon-transmon coupling strength $g_{_{\mathrm{AB}}}=g_{_{\mathrm{BC}}}\approx2\pi\times11.41$ MHz. Suppose the initial state of the three qubits is $\frac{1}{\sqrt2}(|000\rangle+|001\rangle)$. The corresponding ideal final state can be shown as $|\psi_{f_S}\rangle=\frac{1}{\sqrt2}(|000\rangle+|100\rangle)$. Here we evaluate this gate by the state fidelity defined by $F_S=\langle\psi_{f_S}|\rho_2|\psi_{f_S}\rangle$. In order to fully evaluate the performance of the implemented two-qubit gate, we consider a general initial state $|\psi_2\rangle=(\cos\vartheta_1|0\rangle_{_{\mathrm{A}}}+\sin\vartheta_1|1\rangle_{_{\mathrm{A}}})
\otimes|0\rangle_{_{\mathrm{B}}}\otimes(\cos\vartheta_2|0\rangle_{_{\mathrm{C}}}
+\sin\vartheta_2|1\rangle_{_{\mathrm{C}}})$ whose ideal final state takes the form $|\psi_f\rangle=V_S|\psi_2\rangle$. The associated gate fidelity is defined as $F^G_S=\frac {1} {4\pi^2}\int_0^{2\pi} \int_0^{2\pi} \langle \psi_f|\rho_2|\psi_f\rangle d\vartheta_1d\vartheta_2$ with the integration numerically done for 10000 input states with $\vartheta_1$ and $\vartheta_2$ uniformly distributed over $[0, 2\pi]$. In Fig. \ref{Fig3}, we plot the state and gate fidelities as a function of the decoherence rate $\kappa$, represented by blue circles and red triangles, respectively. A gate fidelity as high as $99.41\%$ can be reached when $\kappa=2\pi \times10$ kHz. Note that the best two qubit gate fidelities in the experiment in Ref. \cite{UCSB} is $99.40\%$ for a decoherence rate of around $2\pi \times4$ kHz.

\section{Discussion and Conclusion}

Above we have constructed a set of universal nonadiabatic holonomic quantum gates. To demonstrate the robustness and scalability of our scheme in a more practical scenario, we now perform numerical simulation of a gate sequence which contains both single- and two-qubit gates, and calculate the corresponding fidelity of the whole process. Three qubits (A, B, and E) are involved in the simulation and the gate sequence $U_P^{^{\mathrm{BE}}} V_S^{^{\mathrm{ABE}}} U_P^{^{\mathrm{AB}}}$ is carried out. From right to left, the sequence means applying a $U_P$ on target qubit A with the help of auxiliary qubit B, a SWAP-like gate $V_S$ on qubits A and E, and, finally, a $U_P$ on target qubit E. The parameters of qubit E are $\alpha_{_{\mathrm{E}}}=2\pi\times325$ MHz and $\Delta_3=\omega_{_{\mathrm{B}}}-\omega_{_{\mathrm{E}}}=2\pi\times235$ MHz. The general initial state used in the simulation is $|\psi'\rangle=(\cos\theta'|0\rangle_{_{\mathrm{A}}}+\sin\theta'|1\rangle_{_{\mathrm{A}}})\otimes|00\rangle_{_{\mathrm{BE}}}$.
After application of the gate sequence, the corresponding ideal final state reads $|\psi'_f\rangle=|00\rangle_{_{\mathrm{AB}}}\otimes(\cos\theta'|0\rangle_{_{\mathrm{E}}}+\sin\theta' e^{i\frac {\pi} {2}}|1\rangle_{_{\mathrm{E}}})$. We first choose a specific initial state in which $\theta'=\frac{\pi}{4}$ and plot the state fidelity $F_G=\bra{\psi'_f}\rho'\ket{\psi'_f}$ as a function of the decoherence rate. Then we choose 1001 input states with $\theta'$ uniformly distributed over $[0, 2\pi]$ and obtain the gate fidelity $F_G^G$ by averaging all the state fidelities. The corresponding results are illustrated in Fig.~\ref{Fig4} which shows that the fidelities can reach about $99\%$ within $2\pi \times5$ kHz.

In summary, we have proposed to implement nonadiabatic HQC in a coupled superconducting transmon system. Through control of the amplitudes and relative phases of a driving ac magnetic modulating flux, fast and high-fidelity universal quantum gates on target transmon qubits can be obtained, in a tunable and all-resonant way. Thus, our scheme provides a promising way towards the practical realization of high-fidelity nonadiabatic HQC.

\acknowledgments

This work was supported in part by the National Natural Science Foundation of China (Grant No. 11874156), the National Key R\&D Program
of China (Grant No. 2016YFA0301803), and the China Postdoctoral Science Foundation (Grant No. 2018M631437).


\begin{thebibliography}{99}

\bibitem{shor}
P. W. Shor,
SIAM Rev. \textbf{41}, 303 (1999).

\bibitem{HQC1}
P. Zanardi and M. Rasetti,
Phys. Lett. A \textbf{264}, 94 (1999).

\bibitem{non-Abelian}
F. Wilczek and A. Zee,
Phys. Rev. Lett. \textbf{52}, 2111 (1984).

\bibitem{HQC2}
J. Pachos, P. Zanardi, and M. Rasetti,
Phys. Rev. A \textbf{61}, 010305 (1999).

\bibitem{Duan}
L.-M. Duan, J. I. Cirac, and P. Zoller,
Science \textbf{292}, 1695 (2001).

\bibitem{HQC3}
P. Zhang, Z. D. Wang, J. D. Sun, and C. P. Sun,
Phys. Rev. A \textbf{71}, 042301 (2005).

\bibitem{HQC4}
I. Kamleitner, P. Solinas, C. M\"{u}ller, A. Shnirman, and M. M\"{o}tt\"{o}nen,
Phys. Rev. B \textbf{83}, 214518 (2011).

\bibitem{HQC5}
V. V. Albert, C. Shu, S. Krastanov, C. Shen, R.-B. Liu, Z.-B. Yang, R. J. Schoelkopf, M. Mirrahimi, M. H. Devoret, and L. Jiang,
Phys. Rev. Lett. \textbf{116}, 140502 (2016).

\bibitem{ZSL}
S.-L. Zhu and Z. D. Wang,
Phys. Rev. Lett. \textbf{89}, 097902 (2002).

\bibitem{ZSL1}
S.-L. Zhu and Z. D. Wang,
Phys. Rev. A \textbf{67}, 022319 (2003).

\bibitem{Tong}
E. Sj\"{o}qvist, D. M. Tong, L. M. Andersson, B. Hessmo, M. Johansson, and K. Singh,
New J. Phys. \textbf{14}, 103035 (2012).

\bibitem{XGF}
G. F. Xu, J. Zhang, D. M. Tong, E. Sj\"{o}qvist, and L. C. Kwek,
Phys. Rev. Lett. \textbf{109}, 170501 (2012).

\bibitem{3level1}
J. Zhang, L.-C. Kwek, E. Sj\"{o}qvist, D. M. Tong, and P. Zanardi,
Phys. Rev. A \textbf{89}, 042302 (2014).

\bibitem{3level2}
G.-F. Xu and G.-L. Long,
Sci. Rep. \textbf{4}, 6814 (2014).

\bibitem{3level3}
G. F. Xu, C. L. Liu, P. Z. Zhao, and D. M. Tong,
Phys. Rev. A \textbf{92}, 052302 (2015).

\bibitem{3level4}
Z.-Y. Xue, J. Zhou, and Z. D. Wang,
Phys. Rev. A \textbf{92}, 022320 (2015).

\bibitem{3level5}
Z.-Y. Xue, J. Zhou, Y.-M. Chu, and Y. Hu,
Phys. Rev. A \textbf{94}, 022331 (2016).

\bibitem{3level6}
P. Z. Zhao, G. F. Xu, and D. M. Tong,
Phys. Rev. A \textbf{94}, 062327 (2016).

\bibitem{3level7}
E. Herterich and E. Sj\"{o}qvist,
Phys. Rev. A \textbf{94}, 052310 (2016).

\bibitem{3level8}
G. F. Xu, P. Z. Zhao, T. H. Xing, E. Sj\"{o}qvist, and D. M. Tong,
Phys. Rev. A \textbf{95}, 032311 (2017).

\bibitem{3level9}
Z.-Y. Xue, F.-L. Gu, Z.-P. Hong, Z.-H. Yang, D.-W. Zhang, Y. Hu, and J. Q. You,
Phys. Rev. Appl. \textbf{7}, 054022 (2017).

\bibitem{3level10}
G. F. Xu, P. Z. Zhao, D. M. Tong, and E. Sj\"{o}qvist,
Phys. Rev. A \textbf{95}, 052349 (2017).

\bibitem{3level11}
P. Z. Zhao, G. F. Xu, Q. M. Ding, E. Sj\"{o}qvist, and D. M. Tong,
Phys. Rev. A \textbf{95}, 062310 (2017).

\bibitem{3level12}
Z.-P. Hong, B.-J. Liu, J.-Q. Cai, X.-D. Zhang, Y. Hu, Z. D. Wang, and Z.-Y. Xue,
Phys. Rev. A \textbf{97}, 022332 (2018).

\bibitem{Abdumalikov2013}
A. A. Abdumalikov, J. M. Fink, K. Juliusson, M. Pechal, S. Berger, A. Wallraff, and S. Filipp,
Nature (London) \textbf{496}, 482 (2013). 

\bibitem{ibmexp}
D. J. Egger, M. Ganzhorn, G. Salis, A. Fuhrer, P. Mueller, P. K. Barkoutsos, N. Moll, I. Tavernelli, and S. Filipp,
arXiv: 1804.04900.

\bibitem{xuy}
Y. Xu, W. Cai, Y. Ma, X. Mu, L. Hu, T. Chen, H. Wang, Y. P. Song, Z.-Y. Xue, Z.-q Yin, and L. Sun,
Phys. Rev. Lett. \textbf{121}, 110501 (2018).

\bibitem{sust}
T. Yan, B.-J. Liu, K. Xu, C. Song, S. Liu, Z. Zhang, H. Deng, Z. Yan, H. Rong, M.-H. Yung, Y. Chen, and D. Yu,
arXiv: 1804.08142.

\bibitem{Feng2013}
G. Feng, G. Xu, and G. Long,
Phys. Rev. Lett. \textbf{110}, 190501 (2013).

\bibitem{li2017}
H. Li, L. Yang, and G. Long,
Sci. China: Phys., Mech. Astron. \textbf{60}, 080311 (2017).

\bibitem{Zu2014}
C. Zu, W.-B. Wang, L. He, W.-G. Zhang, C.-Y. Dai, F. Wang, and L.-M. Duan,
Nature (London) \textbf{514}, 72 (2014).   

\bibitem{Arroyo-Camejo2014}
S. Arroyo-Camejo, A. Lazariev, S. W. Hell, and G. Balasubramanian,
Nat. Commun. \textbf{5}, 4870 (2014).

\bibitem{nv2017}
Y. Sekiguchi, N. Niikura, R. Kuroiwa, H. Kano, and H. Kosaka,
Nat. Photon. \textbf{11}, 309 (2017). 

\bibitem{nv20172}
B. B. Zhou, P. C. Jerger, V. O. Shkolnikov, F. J. Heremans, G. Burkard, and D. D. Awschalom,
Phys. Rev. Lett. \textbf{119}, 140503 (2017).

\bibitem{fourlevel1}
V. A. Mousolou, C. M. Canali, and E.~Sj\"{o}qvist,
New J. Phys. \textbf{16}, 013029 (2014).

\bibitem{fourlevel2}
V. A. Mousolou and E. Sj\"{o}qvist,
Phys. Rev. A. \textbf{89}, 022117 (2014).

\bibitem{fourlevel3}
V. A. Mousolou,
Phys. Rev. A \textbf{96}, 012307 (2017).

\bibitem{fourlevel4}
J. Zhang, S. J. Devitt, J. Q. You, and F. Nori,
Phys. Rev. A \textbf{97}, 022335 (2018).

\bibitem{transmon}
J. Koch, T. M. Yu, J. Gambetta, A. A. Houck, D. I. Schuster, J. Majer, A. Blais, M. H. Devoret, S. M. Girvin, and R. J. Schoelkopf,
Phys. Rev. A \textbf{76}, 042319 (2007).

\bibitem{flux-driven1}
J. D. Strand, M. Ware, F. Beaudoin, T. A. Ohki, B. R. Johnson, A. Blais, and B. L. T. Plourde,
Phys. Rev. B \textbf{87}, 220505(R) (2013).

\bibitem{flux-driven0}
Y. X. Liu, C. X. Wang, H. C. Sun, and X. B. Wang,
New J. Phys. \textbf{16}, 015031 (2014).

\bibitem{flux-driven2}
M. Roth, M. Ganzhorn, N. Moll, S. Filipp, G. Salis, and S. Schmidt,
Phys. Rev. A \textbf{96}, 062323 (2017).

\bibitem{flux-driven3}
M. Reagor \emph{et al}.
Sci. Adv. \textbf{4}, eaao3603 (2018).

\bibitem{lix}
X. Li, Y. Ma, J. Han, T. Chen, Y. Xu, W. Cai, H. Wang, Y. P. Song, Z.-Y. Xue, Z.-q. Yin, and L. Sun,
Phys. Rev. Appl. \textbf{10}, 054009 (2018).

\bibitem{OSSE1}
J. Anandan,
Phys. Lett. A \textbf{133}, 171 (1988).

\bibitem{OSSE2}
J. Zhang, T. H. Kyaw, D.M. Tong, E. Sj\"{o}qvist, and L. C. Kwek,
Sci. Rep. \textbf{5}, 18414 (2015).

\bibitem{anharmonicity}
M. H. Devoret and R. J. Schoelkopf,
Science \textbf{339}, 1169 (2013).

\bibitem{decoherence}
C. Rigetti, \emph{et al}., 
Phys. Rev. B \textbf{86}, 100506(R) (2012).

\bibitem{gatefidelity1}
J. F. Poyatos, J. I. Cirac, and P. Zoller,
Phys. Rev. Lett. \textbf{78}, 390 (1997).

\bibitem{SWAPlike}
A. Miranowicz, S. K. \"{O}zdemir, J. Bajer, G. Yusa, N. Imoto, Y. Hirayama, and F. Nori,
Phys. Rev. B \textbf{92}, 075312 (2015).

\bibitem{UCSB}
R. Barends \emph{et al}., 
Nature (London) \textbf{508}, 500 (2014).

\end{thebibliography}
\end{document}